\documentclass[a4paper, 10pt, conference]{ieeeconf}      % Use this line for a4
\IEEEoverridecommandlockouts                              % This command is only
\usepackage[pdftex]{graphicx}
\usepackage{epstopdf}
\usepackage{graphics} % for pdf, bitmapped graphics files
\usepackage{epsfig} % for postscript graphics files
\usepackage{mathptmx} % assumes new font selection scheme installed
\usepackage{times} % assumes new font selection scheme installed
\usepackage{amsmath} % assumes amsmath package installed
\usepackage{amssymb}  % assumes amsmath package installed
\usepackage{subfigure}
\usepackage{url}

\newcommand{\sgn}{\operatorname{sgn}}

\usepackage[utf8]{inputenc}
\title{\LARGE \bf
Data Mining of Causal Relations from Text: Analysing Maritime Accident Investigation Reports
}

\author{Santosh Tirunagari% <-this % stops a space
\thanks{This work is an extract of S. Tirunagari's Master's thesis study submitted at Aalto University School of Science in 2013. He is now a  doctoral student at Department of Computing and CVSSP, University of Surrey, UK.
        {\tt\small s.tirunagari@surrey.ac.uk}}%
}

\begin{document}

\maketitle
\thispagestyle{empty}
\pagestyle{empty}

%%%%%%%%%%%%%%%%%%%%%%%%%%%%%%%%%%%%%%%%%%%%%%%%%%%%%%%%%%%%%%%%%%%%%%%%%%%%%%%%
\begin{abstract}

Text mining is a process of extracting information of interest from text. Such a method includes techniques from various areas such as Information Retrieval (IR), Natural Language Processing (NLP), and Information Extraction (IE). In this study, text mining methods are applied to extract causal relations from maritime accident investigation reports collected from the Marine Accident Investigation Branch (MAIB). These causal relations provide information on various mechanisms behind accidents, including human and organizational factors relating to the accident. The objective of this study is to facilitate the analysis of the maritime accident investigation reports, by means of extracting contributory causes with more feasibility. A careful investigation of contributory causes from the reports provide opportunity to improve safety in future. 

Two methods have been employed in this study to extract the causal relations. They are 1) Pattern classification method and 2) Connectives method. The earlier one uses naïve Bayes and Support Vector Machines (SVM) as classifiers. The latter simply searches for the words connecting cause and effect in sentences. 

The causal patterns extracted using these two methods are compared to the manual (human expert) extraction. The pattern classification method showed a fair and sensible performance with F-measure(average) = 65\% when compared to connectives method with F-measure(average) = 58\%. This study is an evidence, that text mining methods could be employed in extracting causal relations from marine accident investigation reports. 
\end{abstract}

%%%%%%%%%%%%%%%%%%%%%%%%%%%%%%%%%%%%%%%%%%%%%%%%%%%%%%%%%%%%%%%%%%%%%%%%%%%%%%%%
\section{Introduction}
There is a growing concern in the maritime industry regarding human and organizational factors that affect sailing performance and the overall safety of ship operations in and onboard~\cite{bradford2005growing}. This concern stems from a recent rise in commercial maritime accidents caused by ill-fated decisions taken by higher level management. This is further highlighted by academic research showing direct ties between organizational factors and safe performance of maritime crew of the ship. However, effective tools or methodologies for identifying and mitigating potentially harmful human and organizational factors before
they \emph{cause} an accident are yet to be developed. 

The purpose of the present research is to extract the \emph{causal patterns} from accident investigation reports. These patterns study human and organizational factors affecting safety culture and discuss models of safety culture used to design assessment techniques. A careful investigation of these patterns  provides an opportunity to improve and manage safety in the future~\cite{schroder2011accident}. This study aspires to model causal parameters relating accidents. 

\subsection{Motivation}

During the last century, sea trade has been increased due to technological advancements \cite{ladan2012data}. Hence, increasing number of ships are sailing on the world seas. Modern ships are getting faster, bigger and highly automated. Though these technological advancements are beneficial, they still pose a challenge in themselves. Accidents at sea still occur and the consequences to people, ship or environment, are often greater than before~\cite{kujala2009analysis}. 

These accidents are investigated by a maritime accident investigation board. The board reports how the accident occurred, the circumstances, causes, consequences and rescue operations. These reports also provide recommendations for preventing similar accidents. The reports are long, detailed and systematic examinations of marine accidents in order to determine the causes of the accident. 

In this paper, the accident investigation reports are a collection from Maritime Accident Investigation Branch (MAIB). MAIB examines and investigates all types of marine accidents to or on board United Kingdom (UK) ships worldwide, and other ships in UK territorial waters. It includes 11 categories of reports relating to 'Machinery Failures', 'Fire/Explosion', 'Injury/Fatality', 'Grounding', 'Collision/Contact', 'Flooding/Foundering', 'Listing/Capsize', 'Cargo Handling Failure', 'Weather Damage', 'Hull Defects' and 'Hazardous Incidents'. 

Human intervention is required in extracting the causal patterns from the accident investigation reports, as they are in text format. The extraction is generally a difficult job as it takes lot of time and also human may not always be able to extract the interesting information objectively~\cite{ladan2012data}.  Hence, these challenges have been attempted with text mining. As an example, the role of lack of situation awareness in maritime accident causation was examined using a text mining software from accident reports~\cite{grech2002human}. 

\subsection{Previous Studies}

According to~\cite{grech2002human}, causal patterns from the accident investigation reports provide information on
various mechanisms behind accidents. Unfortunately, in the maritime field, no standard reporting formats exist and data collection from the textual reports is a laborious task~\cite{tirunagari2012mining}. Text mining provides a means for efficient and informative scanning of accident cases of interest without reading the actual report. Therefore, text mining in this
context is seen as a useful tool in understanding accidents
and their influencing factors. 

\cite{francis2010text} applied text mining methods on two text
databases, a road accident description and on survey databases. They
extracted new variables from the unstructured text which were later used for
predicting the likelihood of attorney involvement and the severity of
claims. Interesting themes were  identified in the responses
of the survey data. Thus, useful information that would not otherwise
be available was derived from both the databases using text mining
methods. \cite{yiu2011investigation} investigated and validated a novel text mining
methodology for occupational accident analysis and prevention. He also
suggested that adoption of text mining analysis is probably most
feasible for large organizations that can more easily absorb the
labour-intensive steps required to conduct the most meaningful text
mining analysis of occupational injury data. Another article by~\cite{zheng2010analysis} used a text data mining technique called attribute reduction from accident reports to extract most frequent concepts which were considered as the reasons leading to human errors in ship accidents.  An article by~\cite{artana2005development} developed and evaluated software using text mining algorithms for encountering marine hazards. This essential risk management system covered both organizational and human errors.

The previous studies suggest that text mining could be applied on accident
investigation reports. However, application of text mining is a complex
task as it involves dealing with the text data which is
unstructured. Hence, there is an urgent need for a new generation of computational theories and tools to assist
humans in extracting useful information (knowledge) from the rapidly growing volumes
of unstructured accident investigation reports.

\subsection{Research Problem}

Mining the maritime accident investigation reports is a new topic and not much has been covered~\cite{liu2004data}. Until now, it is still regarded as one of the challenging areas since reports have been written in natural language~\cite{tirunagari2012mining}. The latest developments in Natural Language Processing (NLP) and the availability of faster computers facilitates to extract more information from the text. Emphasis should be placed on mining information from unstructured information sources like accident investigation reports.  

The research problem is formulated as follows: 
\begin{itemize}
\item How can causal relations be extracted from maritime accident investigation reports?
\end{itemize}

The following research questions help solving the research problem. 
\begin{itemize}
\item How are the accident investigation reports written and structured?
\item What categories of accident investigation reports should be considered?
\item What models and algorithms should be chosen for this application?
\item How are these models evaluated?
\end{itemize}

These research questions are answered reasonably in this paper. They are intended as a support for solving the research problem. Whilst performing the study, knowledge of classification techniques is also acquired and documented. This section briefly presents the aim, limitations of the study and the structure of paper. 

The main objective is to facilitate analysis of maritime accident investigation reports describing the human and organizational factors in accidents. These factors are extracted as causal relations using text mining methods. The study uses pattern classification and connectives methods to mine causal relations. In both these methods F-measure is used to evaluate the performance. Other rule based techniques including extraction of sentences based on syntactic grammars are left outside the scope of this study. The main reason is that these methods use Parts of Speech (PoS) taggers and  there is no PoS tagger that gives a 100\% accuracy~\cite{manning2011part}. An inaccurate PoS tag can change the grammar of a causal sentence to that of a non-causal.   

\subsection{Limitations} 

The analysis in this study is limited to mining the causal text relating to 'Groundings', 'Collisions', 'Machinery Failures' and 'Fire' related accidents. The scope of the study has also been limited by focusing only on pattern classification and connectives methods for extracting the causal relations to keep the study to a reasonable size.  

There are quite a few challenges when dealing with accident investigation reports. The reports are written in the natural language with no standard template. Misspellings and abbreviations are often found. Detection of compound words such as "safety culture", "spirit status", etc are difficult as order of importance is unknown. The contextual meaning of the words "safety" and
"culture" differs significantly but the word "safety culture" has a different meaning altogether. Therefore, context and semantics play an important role in text mining.

\subsection{Outline}

Section 2 introduces the causal relation extraction methods employed in this study, such as: 1) pattern classification method and 2) connectives method. The former consists of naive Bayes and SVM classifiers and the latter uses connecting words. This chapter also discusses the evaluation techniques such as F-measure, K-fold cross validation and parameter tuning. Section 3 illustrates the data preprocessing techniques such as: tokenization, stop word removal and stemming. It further discusses the document representation. Section 4 presents the experiments and corresponding results. Finally section 5 concludes the paper with discussions.

\section{Methodolgies: Causal Relations Extraction}

A causal relation is the relation between an event (the cause) and a second event (the effect), where the second event is understood as a consequence of the first~\cite{kim1973causes}. In other words, cause is the producer and effect is the result~\cite{hobbs2005toward}.  Causal relations have been studied in several fields.~\cite{white1990ideas} provides an overview of theories within the fields of Philosophy and Psychology. This study explores two different methods for extracting causal relations from maritime accident investigation reports. They are the pattern classification method and connectives method.

\subsection{Pattern Classification Method}
\label{s:intropat}

Pattern recognition is a subfield in machine learning with a purpose of developing methods that recognize meaningful patterns from the data. Pattern recognition has seen applications in the fields of 1) computational fluid dynamics for reduce order modelling~\cite{vuorinen2013large, tirunagari2015exploratory, tirunagarispecial,tirunagari2012analysis}. 2) In forensics, biometrics for detecting spoof images/videos~\cite{tirunagari2015detection, iorliam2015data,tome20151st}. 3) In healtcare applications~\cite{tirunagariidentifying, tirunagari2015breast, poh2014challenges, tirunagari2011s} and 4) in NLP~\cite{tirunagari2012mining, tirunagari2012impact_,paukkeri2011effect,ramaseshan2012twitter, lagus15400natural, puglieseunsupervised}. Pattern classification, on the other hand is a subset of pattern recognition which is based on the classification of features. In other words, pattern classification observes the environment to learn and distinguish patterns of interests and make reasonable decisions about the pattern (or finding the correct class represented by the pattern)~\cite{webb2011statistical}. The decision of the pattern classifiers depend on the prior available patterns. The more relevant patterns are available for the pattern classifier, the better the decision will be. 

\begin{figure}[htbp]

  \centering
  \includegraphics[width=3.5in, height=0.5in]{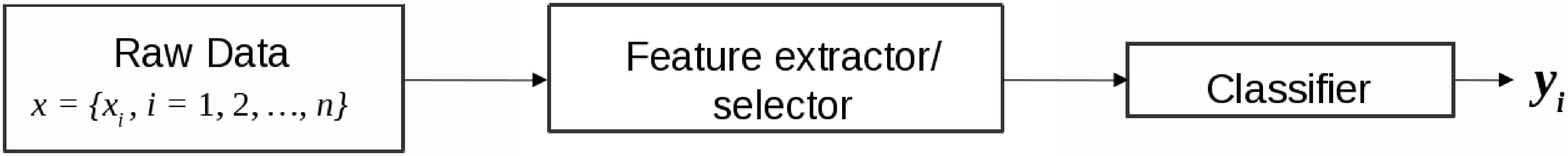}
  \caption[Basic representation of pattern classifier]{\footnotesize Basic representation of pattern classifier.}
  \label{fig:2.7}
\end{figure}

In machine learning, a pattern is a set of attributes that represents a data point $\mathbf{x}$. Let us assume, $\mathbf{x} = (x_1, x_2, ... , x_n)$ to be the pattern, with $x_i, i = \{1, 2, ... , n\}$ being the features of $\mathbf{x}$. Let us assume that these patterns correspond to $P$ number of classes, denoted as $y_i, y_i \in \{1, 2, ... , P\}$ \& $i \in \{1, 2, ..., k\}$ . The graphical representation of a basic pattern classifier is shown in Fig.\ref{fig:2.7}.\\

% figure 2.7 %
% end figure %

Pattern classification methods are of two types, \textit{supervised methods} and \textit{unsupervised methods}. The major difference between supervised and unsupervised methods is the process of learning, during which the characteristics of the data are learned by the classifier. In supervised classification methods, the pattern $\mathbf{x} = (x_1, x_2, ... , x_n)$ along with its associated label or class  $y_i, y_i \in \{1, 2, … , P\}$ \& $i \in \{1, 2, ..., k\}$, form the training dataset $S, \{(\mathbf{x_i}, y_i), i = \{1, 2, ... , k\}\}$. During the training phase, the classifier learns from the existing patterns with their corresponding labels. The trained classifier can then be used to predict the labels for the new unseen data or test data. On the contrary, unsupervised methods do not use labels $y_i$ along with the patterns $\mathbf{x_i}$ during training. The unsupervised methods estimate the hidden patterns in the data to group the given data into several groups or clusters. Hence, unsupervised methods are also referred to as \textit{Clustering Methods}.

This study used two supervised methods, Support Vector Machines (SVM) and naïve Bayes classifiers to classify causal and non-causal patterns. Let $\mathbf{x} = (x_1, x_2, ... , x_n)$ denote a causal or a non-causal pattern, with $x_i, i = \{1, 2, ... , n\}$ being the Bag of Words (Bow) of $k$ patterns $\mathbf{x_i}, i = \{1, 2, ... , k\}$. These patterns correspond to $2$ number of classes, denoted as $y_i \in \{-1, +1\}$. 

In the following sub-sections, the classifiers and their evaluation techniques are discussed. The figures in the section~\ref{ss:svm} are adapted from "Learning with Kernels"~\cite{scholkopf2002learning} and "kernel methods for pattern analysis"~\cite{shawe2004kernel}.

%% explain how it is related to this study

\subsubsection{Support Vector Machines (SVM)}
\label{ss:svm}
Kernel
Support Vector Machine (SVM) is a widely used pattern classification method and is well known for accurate and effective pattern classification \cite{mitchell1997artificial, vapnik1998statistical, vapnik2000nature}. 

Let $(\mathbf{X,Y}), \mathbf{X} \subseteq R_n ; \mathbf{Y} \in \{-1, +1\}$ denote training data $S$ in a two-class classification task. Each point $\mathbf{x} \in \mathbf{X}$ is associated with one of the possible classes $\mathbf{Y} \in \{-1, +1\}$. The goal of the SVM is to classify a new data point $\mathbf{x'}$ to one of the possible classes. In probabilistic notation, the likelihood that a new point $\mathbf{x'}$ belongs to a given class, $y’ \in \{+1, -1\}$, can be represented as, 
\begin{eqnarray}
	p(y' = +1| \mathbf{x'} = \mathbf{x}), \nonumber \\
	p(y' = -1| \mathbf{x'} = \mathbf{x}). \nonumber
\end{eqnarray}
					
Now, the classifier $f : \mathbf{X} \rightarrow \mathbf{Y}$ estimates the representation of the discriminant function. During training, the function $f$ has to minimize the probability of misclassification of all data points in the training data. 

SVM solves this problem by finding the function $f$, which for every point $(\mathbf{x_i}, y_i); \mathbf{x_i}= [x_{i1}, x_{i2}, … , x_{in}]^{T} \in \mathbf{X}, y_i \in \{-1, +1\}$, in the training set satisfies,
\begin{eqnarray}
\label{eq:1}
	f(\mathbf{x_i}) \geq 0, \text{if } y_i = +1, \nonumber \\
	f(\mathbf{x_i}) < 0, \text{if } y_i = -1.
\end{eqnarray}

Eq.(\ref{eq:1}) is only possible if there exists a hypersurface $h$, which can separate the data into two classes either linearly or non-linearly. 

% Figure 2.8 %
\begin{figure}

  \centering
  \includegraphics[width=0.7\linewidth]{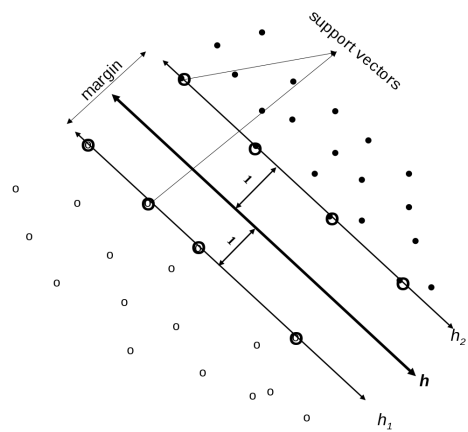}
  \caption[Binary SVM in linearly separable case]{\footnotesize The optimal separating hyperplane $\mathbf{h}$ in linearly separable binary classification using support vector machine (SVM). Support vectors are shown in the highlighted circles that lie on the hyperplanes (dotted lines, $h_1$ and $h_2$) that have unit distance to the optimal separating hyperplane (solid line, $\mathbf{h}$). }
  \label{fig:2.8}
\end{figure}
% end figure %

\textbf{Linearly separable Binary classification (maximal margin)}\\
Let us assume that we have a linearly separable training data set, $\mathbf{S} = \{(\mathbf{x_i}, y_i)\}$, $ i=1,2,...,k$, where $\mathbf{x_i}$ is any single data point and $y_i$ is the corresponding class label of $\mathbf{x_i}$ and there are $k$ data points in $\mathbf{S}$. The decision function $\sgn(g(\mathbf{x}))$ is equal to the sign of the $g(\mathbf{x})$, where $g(\mathbf{x})$ is any function of $\mathbf{x}$.
\begin{eqnarray}
      \sgn(g(\mathbf{x})) =  \begin{cases}
                 +1, g(\mathbf{x}) \geq 0,\\
		 -1, g(\mathbf{x}) < 0
                \end{cases}
\end{eqnarray}

For the given set of training data $\mathbf{S}$, there exists a linear discriminant function $f$ of the form, 
\begin{eqnarray}
    f : \mathbf{x} \rightarrow \mathbf{w}^{T}\mathbf{x}+b, \nonumber
\end{eqnarray}

where, $\mathbf{w} \in \mathbb{R}^{n}$, $b \in \mathbb{R}$ is a constant and the corresponding decision function, $t = \sgn(\mathbf{w}^{T}\mathbf{x}+b)$ should have zero error. This means, all the $k$ data points in the training data set, $\mathbf{S}$ should satisfy the decision function $t$. So, it is possible that there exists infinite number of such hyper-planes ($h$) that can separate the two classes with zero error. The goal of SVM is to maximize the minimal distance between the two hyperplanes ($h_1$ and $h_2$) that can separate the data (minimal margin, as shown in Fig. \ref{fig:2.8}) of the linear discriminant function $f$ with respect to the training data set $\mathbf{S}$ \cite{kressel1999pairwise}.
\begin{eqnarray}
    min_{\mathbf{x_i} \in \mathbf{X}}|\mathbf{w}^{T} \mathbf{x_i} + b|. \nonumber
\end{eqnarray}

The geometric margin $\gamma$ for the discriminant function is defined as, 
\begin{eqnarray}
\label{eq:2}
	\gamma = \frac{1}{\|\mathbf{w}\|}%\gamma_i  = \frac{y_i (w^{T} x_i + b)}{\|w\|^{2}}.
\end{eqnarray}

From Eq.(\ref{eq:2}), it is clear that maximizing the minimal geometric margin reduces to minimizing the norm of the weight vector, $\|\mathbf{w}\|^{2}$. The hyperplane that maximizes the minimum margin and satisfies
\begin{eqnarray}
\label{eq:3}
	y_i(\mathbf{w}^{T} \mathbf{x_i} + b) \geq 1, i = \{1, 2, ... , k\},
\end{eqnarray}

is called the \textit{optimal separating hyperplane} \cite{kressel1999pairwise}.

From Eq.(\ref{eq:2}) and Eq.(\ref{eq:3}) the optimal separating hyperplane can be represented as follows,
\begin{eqnarray}
	min_w & \frac{1}{2}\|\mathbf{w}\|^{2}_2, \nonumber
\end{eqnarray}
\begin{eqnarray}
\label{eq:4}
	\text{such that }, y_i(\mathbf{w}^{T} \mathbf{x_i} + b) \geq 1, i = \{1, 2, ... , k\}.
\end{eqnarray}
% figure 2.9 %
\begin{figure}[!t]

  \centering
  \includegraphics[width=0.7\linewidth]{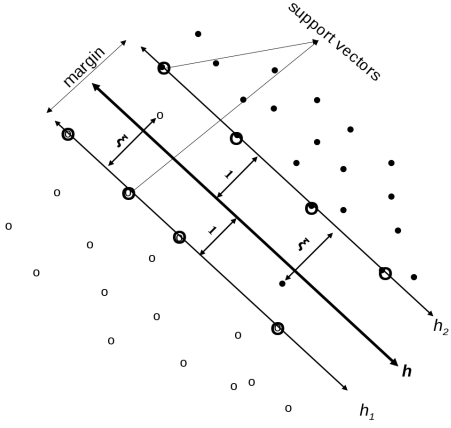}
  \caption[Binary SVM in linearly non-separable case]{\footnotesize The optimal separating hyperplane, $\mathbf{h}$ in the linearly non-separable binary classification using support vector machine (SVM). Support vectors are shown in the highlighted circles that lie on the hyperplanes (dotted lines, $h_1$ and $h_2$) that have unit distance to the optimal separating hyperplane (solid line, $\mathbf{h}$). }
  \label{fig:2.9}
\end{figure}
% end figure %
The above minimization problem can be solved as a dual optimization problem using \textit{Lagrangian}, $\mathbb{L}$.
\begin{eqnarray}
\label{eq:5}
	\mathbb{L} = min_w max_\alpha \{\frac{1}{2} \|\mathbf{w}\|^{2} - \sum^{k}_{i=1} \alpha_i[y_i(\mathbf{w}^{T}\mathbf{x} – b) – 1]\},
\end{eqnarray}
where $\alpha_i$ is the lagrangian multiplier. 

The solution of the above optimization problem Eq.(\ref{eq:5}) defines a linear optimal separating hyperplane defined by the parameters,
\begin{eqnarray}
    \hat{\mathbf{w}} = \sum\limits_{i:\hat{\alpha} > 0} \hat{\alpha}_i \mathbf{x}_i y_i, \nonumber
\end{eqnarray}
\begin{equation}
    \hat{b} = -\frac{1}{2}[\min_{y_i=1} (\mathbf{\hat{w}}^{T} \mathbf{x}_i) + \max_{y_i=-1}(\mathbf{\hat{w}}^{T} \mathbf{x}_i)]. \nonumber
\end{equation}

Training vectors $\mathbf{x_i}$, for which $\alpha_i$ are strictly positive are called support vectors. These support vectors lie on hyperplanes at unit distance from the optimal separating hyperplane (as shown in Fig.\ref{fig:2.8}). Using the above optimized $\hat{w}$ and $\hat{b}$, the classifier $t$ is defined as, 
\begin{eqnarray}
	t(\mathbf{x}) = \sgn(\hat{\mathbf{w}}^{T}\mathbf{x} + \hat{b}). \nonumber
\end{eqnarray}

\textbf{Linearly non-separable Binary classification (Soft-margin)}\\
The perfect linear separability is not realistic. Therefore, we still need to solve the optimization problem to find optimal linear discriminant function. Allowing a certain amount of misclassification, and punishing the misclassified data points during the optimization helps us to resolve linear separability. The amount by which the discriminant function fails to reach the unit margin is termed as \textit{the error of observation}, $\xi$ (as shown in Fig.\ref{fig:2.9}). 
\begin{eqnarray}
	\xi_i = max\{0, 1 – y_i(\hat{\mathbf{w}}^{T}\mathbf{x_i} + \hat{b})\}. \nonumber
\end{eqnarray}
The misclassification takes place when $\xi_i > 1$. \\

For the linearly non-separable data, the optimal separability hyperplane has to maximize the geometric margin and minimize the error function $\Theta(\xi)$. 
\begin{eqnarray}
	\Theta(\xi) = \sum^k_{i=1} \xi_i. \nonumber
\end{eqnarray}
Considering the error $\xi$, the constraint Eq.(\ref{eq:3}) can be written as, 
\begin{eqnarray}
	y_i(\mathbf{w}^{T}\mathbf{x_i} + b) \geq 1-\xi_i , i = \{1, 2, ... , k\} \text{ and } \xi_i \geq 0.	\nonumber
\end{eqnarray}

Now the optimization problem (Eq.\ref{eq:4}) can be written as,
\begin{eqnarray}
\label{eq:6}
	& min_{\mathbf{w},b,\xi} (1/2)\| \mathbf{w} \|^{2}_2 + C \sum^{k}_{i=1} \xi_i, \nonumber \\
	\text{such that }, & y_i(\mathbf{w}^{T}x_i + b) \geq 1 – \xi_i, \text{and} \xi_i \geq 0 ; i = \{1, 2, … , k\},
\end{eqnarray}
where $C$ is a positive parameter, which defines the importance of misclassification errors. \\

To solve the optimization problem of constraint (Eq.\ref{eq:6}), we consider solving the corresponding dual problem with the objective function to be maximized, 
\begin{eqnarray}
\label{eq:7}
	W(\alpha) = \sum^{k}_{i=1} \alpha_i – \frac{1}{2} \sum^{k}_{i,j=1}  \alpha_i\alpha_j y_iy_j (\mathbf{x_i}^{T}\mathbf{x_j}),	\nonumber \\
	 0 \leq \alpha_i \leq C, i = \{1, 2, ... , k\}; \text{and}  \sum^{k}_{i=1}\alpha_iy_i = 0.
\end{eqnarray}

From the constraint Eq.(\ref{eq:7}), $\alpha_i = C$ if and only if $\xi_i > 0$, and the vectors $x_i$ with $\xi_i > 0$ are called \textit{support vectors}.

% figure 2.10 %
\begin{figure}[!t]

  \centering
  \includegraphics[width=0.9\linewidth]{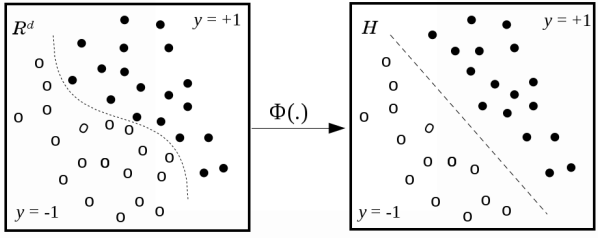}
  \caption[SVM kernel trick]{\footnotesize A non-linear SVM can be interpreted as a linear SVM in a non-linearly mapped space. $\Phi(.)$ defines the non-linear mapping of data from a lower dimension to a higher dimension.}
  \label{fig:2.10}
\end{figure}
% end figure %

\textbf{``Kernel trick" for Non-linear SVM''} \\
The '\textit{kernel trick}' in the context of SVM is a non-linear transformation to map the data in the low dimensional space onto a higher dimensional space (as shown in Fig.\ref{fig:2.10}). By non-linearly mapping the data onto the higher dimensional space with an appropriate kernel, it is supposed that the original linearly non-separable data becomes linearly separable~\cite{scholkopf2002learning, shawe2004kernel}. Since SVM learning needs only the inner product between data points, the non-linear transformation does not apply to individual points in the training set, there by maintaining the efficiency of SVM using the kernel trick. The kernel based SVM often outperforms the original SVM for linearly non-separable classification tasks~\cite{cristianini2000introduction}. The standard kernels are as follows: 

\begin{itemize}
  \item Linear Kernel
    \begin{equation}
      \mathrm{K}(\mathbf{x_i},\mathbf{x_j})=<\mathbf{x_i},\mathbf{x_j}>.
    \end{equation}
 
  \item Gaussian Kernel
    \begin{equation}
      \mathrm{K}(\mathbf{x_i},\mathbf{x_j})=\exp{\left(-\frac{\|\mathbf{x_i}-\mathbf{x_j}\|^2}{\sigma}\right)}.
    \end{equation}
  \item Polynomial Kernel
    \begin{equation}
      \mathrm{K}(\mathbf{x_i},\mathbf{x_j})=(<\mathbf{x_i},\mathbf{x_j}>+\,c)^k.
    \end{equation}

 \end{itemize}

\subsubsection{Naïve Bayes classification}
\label{ss:naivebayes}

The naïve Bayes Classifier is a supervised learning method based on Bayes Rule of probability \cite{mitchell1997artificial}. Naïve Bayes classification algorithms are currently some of the most used pattern recognition algorithms. It is popular for its quick training speeds and high accuracies \cite{mitchell1997artificial, mackay2003information, bishop2006pattern}. 

According to Bayes rule, the posterior belief $P(y|\mathbf{x})$ is calculated by multiplying the prior $P(y)$ by the likelihood $P(\mathbf{x}|y)$ that $\mathbf{x}$ will occur if and only if $y$ is true. Bayes rule is given by,
\begin{eqnarray}
\label{eq:2.8}
	P(y|\mathbf{x}) = \frac{P(\mathbf{x}|y) P(y)}{P(\mathbf{x})}.
\end{eqnarray}

Consider a supervised learning problem, $f : \mathbf{x} \rightarrow y$. To learn $P(y|\mathbf{x})$, we need to approximate the target function $f$. Let us assume, $\mathbf{x} = (x_1, x_2, ... , x_n)$, where $x_i$ is a Boolean random variable denoting the $i^{th}$  attribute of $\mathbf{x}$ and $y$ is a Boolean valued random variable. Applying Bayes rule Eq.(\ref{eq:2.8}) to $P(y = y_i| \mathbf{x} = \mathbf{x_k})$ can be represented as, 
\begin{eqnarray}
	P(y = y_i | \mathbf{x} = \mathbf{x_k}) =\frac{P(\mathbf{x} = \mathbf{x_k} | y = y_i) P(y = y_i)}{\sum_j P(\mathbf{x} = \mathbf{x_k} | y = y_j) P(y = y_j)},
\end{eqnarray}

where $y_i$ is the $i^{th}$ possible value of $y$, $\mathbf{x_k}$ is the $k^{th}$ possible vector of $\mathbf{x}$. 

During learning $P(\mathbf{x}|y)$ and $P(y)$ can be estimated using the training data. Using these estimates, together with Bayes' rule in Eq.(\ref{eq:2.8}), we can determine $P(y|\mathbf{x} = \mathbf{x_k})$ for any new data point $\mathbf{x_k}$. Bayesian classifiers are computationally very expensive; however the Conditional Independence assumption of naïve Bayes algorithm drastically reduces the number of parameters to be estimated when modeling $P(\mathbf{x_k}|y)$, from $2(2n – 1)$ to $2n$. 

\textit{Conditional Independence}: Given random variables $x$, $y$, and $z$; $x$ can be called conditionally independent of $y$ given $z$, if and only if, the probability distribution of $x$ is independent of the value $y$ given $z$.
\begin{eqnarray}
\label{eq:9}
	(\forall_{i,j,k}) P(x &=& x_i | y = y_j, z = z_k) = P(x = x_i | z = z_k).
\end{eqnarray}

The naïve Bayes algorithm assumes the attributes $x_1, x_2, ..., x_n$ which are all conditionally independent of one another given the class $y$. Considering the Conditional Independence assumption of naïve Bayes, we have
\begin{eqnarray}
\label{eq:10}
	P(x_1, x_2, ..., x_n | y) =  \prod^{n}_{i=1} P(x_i | y).
\end{eqnarray}

Now, using Bayes rule and the conditional independence property (Eq.\ref{eq:9}), the probability that $y$ takes the $k^{th}$ possible value given $\mathbf{x}$ is given by,
\begin{eqnarray}
\label{eq:11}
	P(y = y_k |x_1, x_2, ...,x_n) = \frac{P(y = y_k) \prod_i P(x_i|y=y_k)}{\sum_j P(y=y_j)\prod_i P(x_i|y=y_j)}.
\end{eqnarray}

During the training, the distributions $P(y)$ and $P(x_i| y)$ are estimated. Given the attributes of $\mathbf{x'}$ (a new data point), the most probable value of $y$ given $\mathbf{x'}$ can be estimated as, 
\begin{eqnarray}
\label{eq:12}
	y' \leftarrow \text{argmax}_{y_k} P(y = y_k) \prod_i P(x_i | y = y_k).
\end{eqnarray}

\subsubsection{Evaluation: pattern classification method} 

Machine learning algorithms induce classifiers that depend on the training set. So there is a need for evaluation and statistical testing to assess the expected error rate of a classification algorithm. Additionally evaluation is crucial to compare the expected error rates of two classification algorithms to identify the better performing one. Evaluation can also be used as a guide for future improvements on the model. The technique here is to generate a test-set, whose labels are already known. This test-set has to be distinct from the train-set which has been used to train the classifier. The test-set is then labelled by the classifier and the labels that it decides are being compared with their correct labels. 

Additional techniques have been implemented in order to get more accurate evaluations and avoid possible 'over-fitting'. There is a chance that the classifier will become more accurate in the train set and less accurate in the test set with some parameter changes. This is when over-fitting occurs to the train set. 

\textit{k-Fold Cross Validation} 
Cross-validation is a method of evaluating learning algorithms by segmenting the data into several folds, where the folds 
are either training or validation sets.  Each training set is used to
train a model while the validation set is used for validating the
performance of the trained model.  Performance is measured as accuracy
averaged over all folds. 

\begin{figure}

  \centering
  \includegraphics[width=0.6\linewidth]{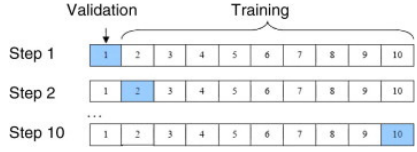}
  \caption[10-fold cross-validation procedure. ]{\footnotesize 10-fold cross-validation procedure. The light-blue folds
    represent the validation folds, while the remaining represent the
    training folds. }
  \label{fig:3fold_cv}
\end{figure}

The most basic form of cross-validation is \textit{k}-fold
cross-validation~\cite{bishop2006pattern}, where the data is first partitioned into \textit{k}
folds of equal or nearly equal size.  Subsequently \textit{k}
iterations of training and validation are performed, such that for
each iteration, the model is validated against a different fold and
trained on the \textit{k}$-1$ folds, as illustrated in Figure~\ref{fig:3fold_cv}. 

% Figure 2.10 %
% end figure %

The next step is to determine the suitable value of \textit{k}. Large $k$ is desirable since it yields more performance estimates.  However it also yields a
lower validation set size, leading to less precise measurements of the
performance metric.  In data mining community, there is general
consensus that \textit{k}$=10$ is a good compromise of these factors,
where making predictions using 90\% of the data makes it more likely
to be generalized to the full data~\cite{kohavi1995study}. 

The results of cross-validation can yield misleadingly low error estimates. The detailed discussion of pitfalls in connection with cross-validation is found in~\cite{forman2010apples}. In this study \textit{k}$=10$ is used.

\textit{Performance Measurements} 
Consider a binary classifier (a predictor) that classifies each pattern in a data set into two classes, either positive (P') or negative (N'), while the ground truth is either positive (P) or negative (N). 
The performance of the classifier can be represented in terms of these four  possible classification results:

True positive (TP): the result is positive (P') while the ground truth is also positive (P)\\
 False positive (FP): the result is positive (P') but the ground truth   is negative (N)\\
 True negative (TN): the result is negative (N') while the ground truth   is also negative (N)\\
False negative (FN): the result is negative (N') but the ground truth   is positive (P)\\

All such symbols can be also treated as the number of patterns that belong to
each of the cases, and we have
\[ \left\{\begin{array}{l} P'=TP+FP \\ N'=TN+FN \end{array} \right.,\;\;\;\;\;\;\;
\left\{\begin{array}{l} P=TP+FN \\ N=TN+FP \end{array} \right. \]

\begin{figure}[h]

  \centering
  \includegraphics[width=0.6\linewidth]{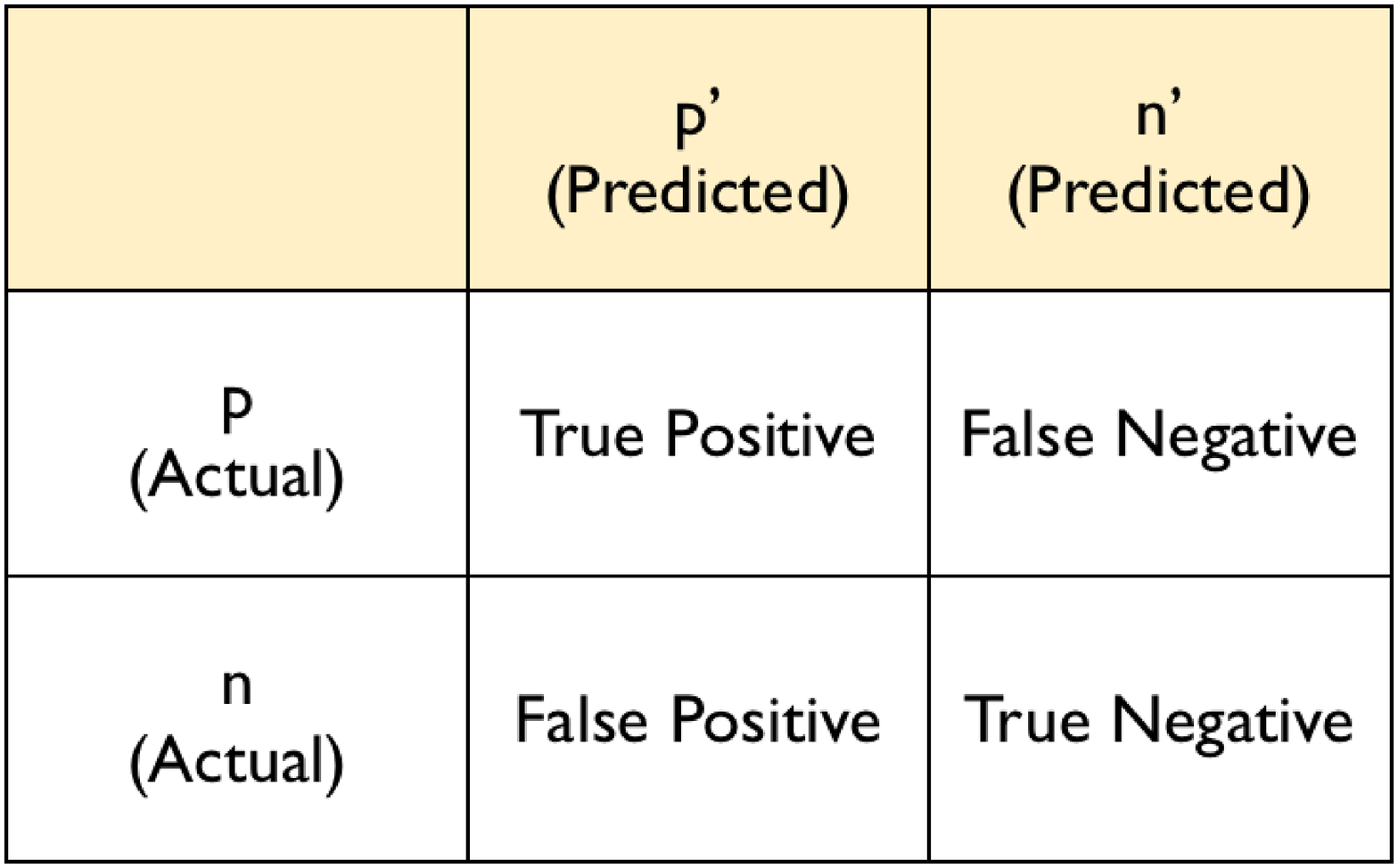}
  \caption[A simple confusion matrix]{A simple confusion matrix.}
  \label{fig:confisionMatixCalc}
\end{figure}

The four cases of the classification result can be represented by the following
2 by 2 confusion matrix (see Figure~\ref{fig:confisionMatixCalc}). Each column of the matrix represents the instances in a predicted class, while each row represents the instances in an actual class. Thus, the diagonal entries indicate labels that were correctly predicted, and the off-diagonal entries indicate errors. One benefit of a confusion matrix is that it is easy to see if the system is confusing two classes. 

% Figure 2.11 %
% end figure %

Based on these concepts, we can further define the following performance measurements (all in percentage between 0 and 1). Sensitivity and specificity are statistical measures of the performance of a binary classification. Sensitivity (true positive rate, or recall) measures the proportion of actual positives which are correctly identified. Specificity measures the proportion of negatives which are correctly identified. An ideal classifier should have $100\%$ sensitivity and $100\%$ specificity. 

The recall and the precision can be derived from the confusion matrix by applying the formulas from the  Table~\ref{tab:met.5}. Recall describes the completeness of the classification. Precision defines the actual accuracy of the classification. 

\begin{table}[h]
\centering
\begin{tabular}{|p{3cm}|c|}
\hline
\textbf{Method} & \textbf{Formula}\\
\hline
 \footnotesize Accuracy &  \footnotesize $ACC=\frac{TP+TN}{P+N}$\\\hline
\footnotesize Error rate &    \footnotesize $ERR=\frac{FP+FN}{P+N}=1-ACC$\\\hline
\footnotesize Recall or True positive rate or Sensitivity & \footnotesize $TPR$ or $Re =\frac{TP}{P}=\frac{TP}{TP+FN}$\\\hline
\footnotesize Precision  & \footnotesize $Pr=\frac{TP}{P'}=\frac{TP}{TP+FP}$\\\hline
\footnotesize True negative rate or Specificity &  \footnotesize $TNR=\frac{TN}{N}=\frac{TN}{FP+TN}$\\\hline
\footnotesize False positive rate & \footnotesize $FPR=\frac{FP}{N}=\frac{FP}{FP+TN}=1-\frac{TN}{N}=1-TNR$\\\hline
\footnotesize False negative rate & \footnotesize  $FNR=\frac{FN}{P}=\frac{FN}{TP+FN}=1-\frac{TP}{P}=1-TPR$\\\hline
\end{tabular} 
\caption{Performance measurement methods}
\label{tab:met.5}
\end{table}

While recall and precision rates can be individually used to determine the quality of a classifier, it is often more convenient to have a single measure to do the same assessment. The F-measure combines the recall and precision rates in a single equation:

\[ F = 2*\frac{precision*recall}{precision+recall} \] 

\textit{F-measure for Cross Validation}
In the previous subsection, the general formula for calculating the F-measure was discussed. ~\cite{forman2010apples} gave a description of three different combination strategies for cross-validation which allow different ways of handling F-measure,  one of them being unbiased.  

The first combination starts with simply averaging of F-measures. In each fold the F-measure is recorded as $F^{(i)}$ and the final estimate is calculated as the mean of all folds. 

\[ F_{avg} := \frac{1}{k} \displaystyle\sum\limits_{i=1}^k  F^{(i)}\] 

The second combination considers,  averaging precision and recall across all the folds. Hence, the final estimate of F-measure can be given as follows: 

\[ Pr := \frac{1}{k} \displaystyle\sum\limits_{i=1}^k  Pr^{(i)} \]
\[ Re := \frac{1}{k} \displaystyle\sum\limits_{i=1}^k  Re^{(i)} \]
\[ F_{pr,re} := 2*\frac{Pr*Re}{Pr+Re} \] 

The third and final combination considers averaging of true positives and false positives across all the folds. This combination is also considered to be unbiased according to the authors. 

\[ TP := \frac{1}{k} \displaystyle\sum\limits_{i=1}^k  TP^{(i)}\]
\[ FP := \frac{1}{k} \displaystyle\sum\limits_{i=1}^k  FP^{(i)}\]
\[ FN := \frac{1}{k} \displaystyle\sum\limits_{i=1}^k  FN^{(i)}\]
\[ F_{tp,fp} := \frac{(2*TP)}{2* TP + FP + FN} \] 

On the evidence provided by the article~\cite{forman2010apples}, this study used \emph{unbiased F-measure ($F_{tp,fp}$)} to evaluate the performance of the \textit{K}-fold cross validation.

\textit{Choice of Parameters}: Most supervised learning algorithms include one or more configurable parameters. The problem is to identify the suitable values for these parameters. Generally, a finite set is defined with alternative values for each parameter. Then, the simplest approach is to run the algorithm with the same training data for each combination of parameter values and measure performance each time on the same validation set~\cite{hsu2003practical}. The parameters that give the best performance on validation set are chosen.

\subsection{Connectives Method}
\label{s:conmen}

The words which are used to connect the cause and effect in sentences are called connecting words. There are approximately  a list of 200 commonly used English connecting words~\cite{transitionwords}. These words introduce a certain shift in the line of argument. Connectives method involves
extracting the causal sentences using these connecting words. These connecting words are usually a transition
or a conjunction~\cite{huddleston2002cambridge, quirk1985comprehensive} or a verb phrase~\cite{girju2002text}. The examples in this chapter are taken from the grammar-quizzes\footnote{\url{http://www.grammar-quizzes.com/19-2.html}}
website.

\subsubsection{Transitions}

Transitions are phrases or words used to connect one idea to the next~\cite{goldman1972pauses}. They may be "Additive",  "Adversative", "Causal",  or "Sequential"~\cite{winterowd1970grammar}. This study considers transition words as words which after a particular time, show a consequence or an effect. More detailed information regarding the transition words can be found in~\cite{transitionwords}. Table~\ref{tab:1} shows the terms which serve as a transition from one sentence to the next. 
\begin{table}
\centering
\begin{tabular}{  p{3cm}  l  p{3cm}}
CAUSE (REASON) & TRANSITION & EFFECT (RESULT) \\ \hline
 She had no other options. & Consequently, & she married at thirteen. \\ \hline
 She was not protected. & As a result, &  she had a baby at thirteen. \\ \hline
 She had no access to health education or medical clinics. & Therefore,&  she was more likely to get HIV.\\ \hline
 There was poor sanitation in the village. & As a consequence, &  she had health problems.\\ \hline
 The water was impure in her village. & For this reason, & she suffered from parasites.\\ \hline
 She had no shoes, warm clothes or blankets. & For all these reasons,&  she was often cold.\\ \hline
 She had no resources to grow food.(land, seeds,tools) & Thus, & she was hungry.\\ \hline
 She had not been given a chance, & so &  she was fighting for survival.\\ \hline
\end{tabular}
\caption{Cause (reason) and effect (result) with transition.}
\label{tab:1}
\end{table}

\subsubsection{Conjunctions}

Conjunctions are the connecting words that are often used to join two complete sentences. The conjunctions, that are used to connect the cause and effect sentences are 'because', 'as', 'since' and 'so'. 'Because', 'as', and 'since' introduce a cause and 'so' introduces an effect. Hence these are used to join two independent clauses together~\cite{winterowd1970grammar}. As shown in Table~\ref{tab:2}, 'because' and other conjunctions, join one clause with another clause. Conjunction introduces a cause (reason) for the situation stated in the other clause.

\begin{table}
\begin{tabular}{  p{3cm}  l  p{3cm}}
\hline
 EFFECT (RESULT) &  CONJUNCTION & CAUSE (REASON)\\\hline
 She married at thirteen & because &  she had no other options.\\\hline
 She had a baby at thirteen & as &  she was not protected.~\\\hline
 She was more likely to get HIV & since &  she had no access to health education. \\\hline
 She had health problems & because of &  poor sanitation in the village.\\\hline
 She suffered from parasites &  on account of &  the impure water in her village.\\\hline
 She was often cold & due to &  not having shoes, warm clothes or blankets.\\\hline
 She was hungry & for the reason that &  she had no resources to grow food.\\\hline
 She was fighting for survival & since &  she had not been given a chance.\\\hline
 \end{tabular}
\caption{Effect (result) and cause (reason) with conjunction.}
\label{tab:2}
\end{table}

\subsubsection{Verb Phrases}

Verb phrases are the part of a sentence containing the verb and a object~\cite{winterowd1970grammar}. These can be used as connecting words to join two noun phrases i.e $<$Noun Phrase 1$><$Verb Phrase$><$Noun Phrase 2$>$. This syntactic structure serves as a causal relation, where the verb phrase acts as a causal verb or reflects a resulting effect in the object. 

Table~\ref{tab:3} shows causal relations with verb phrases. Here the verb phrase introduces the effect in the cause and result expressions. Both verbs "cause" and "result" are used in the active form.

\begin{table}[h]
\begin{tabular}{  p{2cm}  l  p{2cm}}
 CAUSE (REASON) &  VERB PHRASE & EFFECT (RESULT)\\\hline
 Poor childhood education & causes &  illiteracy.\\\hline
 Poor childhood education & results &  in illiteracy.\\\hline
\end{tabular}
\caption{Cause (reason) and effect (result) with verb phrases.}
\label{tab:3}
\end{table}

In Table~\ref{tab:4}, both verbs "cause" and "result" are used to introduce a cause. The verb cause may be used in the passive form with a "by phrase".  The verb result does not take the passive form.  Instead, it is followed by a prepositional phrase "from".

\begin{table}[h]
\begin{tabular}{  p{2cm}  l  p{2cm}}
 EFFECT (RESULT) &  VERB PHRASE &  CAUSE (REASON)\\\hline
 Illiteracy  is & caused &  by poor childhood education.\\\hline
 Illiteracy & results / is resulted by  &  from poor childhood education. \\\hline
\end{tabular}
\caption{Effect (result) and cause (reason) with verb phrases.}
\label{tab:4}
\end{table}

\cite{girju2002text} extracted causal relations which included this syntactic structure. Using this method, they achieved approximately 66\% recall on a test corpus generated from an archive of Los Angeles Times articles. They classified the verb phrases present in causal relations in to four categories: 

\begin{itemize}
\item Low ambiguity and high frequency (LAHF).
\item Low ambiguity and low frequency (LALF). 
\item High ambiguity and low frequency (HALF). 
\item High ambiguity and high frequency (HAHF).
\end{itemize}
The verb phrases which have LAHF are as follows: "cause", "affect", "induce", "produce", "generate", "affect", "arouse", "elicit", "lead to", "trigger", "derive", "associate", "relate to", "link", "originate", "bring on", and "result". 

This study concentrates only on verb phrases such as "cause" and "result", since they have no ambiguity. 

\subsubsection{Evaluation for connectives method}

In the context of connectives method, precision and recall are defined in terms of a set of retrieved causal sentences (e.g. all the causal sentences marked by the automatic algorithm (A)) and a set of relevant causal sentences (e.g. the causal sentences that marked by expert only (E)). 

In here, precision is the fraction of retrieved causal sentences that are relevant to the expert. And recall is the fraction of expert marked causal sentences that are successfully retrieved. It is trivial to achieve recall of 100\% since causal sentences marked by expert and algorithm are not always the same. Therefore, recall alone is not enough but one needs to measure the number of non-relevant causal sentences according to expert. These two measures are used together in the F-measure to provide a single measurement for a system. 

Retrieved := Algorithm marked causal sentences (A). 
Relevant := Expert marked causal sentences (E). 

\[ Precision := \frac{(E \cap A)}{A} \]
\[ Recall := \frac{(E \cap A)}{E} \]

\section{Data, Processing \& Representation}
\label{s:textpre}
The data used in the study is 'MAIB accident investigation reports'. Marine Accident Investigation Branch (MAIB~\footnote{\url{http://www.maib.gov.uk/home/index.cfm}}) is a branch of the Department for Transport located in Southampton, UK. MAIB has four teams of experienced accident investigators, each comprising a principal inspector and three inspectors drawn from the nautical, engineering, naval architecture or fishing disciplines. The role of the MAIB is to contribute to safety at sea by determining the causes and circumstances of marine accidents and working with others to reduce the likelihood of such accidents recurring in the future~\cite{branch2004bridge}. 

There are 11 categories of accident investigation reports which are Machinery, Fire/Explosion, Injury/Fatality, Grounding, Collision/Contact, Flooding/Foundering, Listing/Capsize, Cargo Handling Failure, Weather Damage, Hull Defects and Hazardous Incidents. But this study concentrates only on 4 types of accident types with a total of 135 investigation reports as shown in the Table~\ref{tab:3.1}. Each report, on an average contains 60 pages which are divided into 3 sections viz: 1) narrative 2) analysis and 3) conclusions.  Narrative section describes the summary of the accident, while every possible detail regarding the accident is analyzed in the analysis section.

\begin{table}
\begin{center}
\begin{tabular}{|c|c|}
\hline
 \textbf{Accident Type} &  \textbf{Documents}\\\hline
 Collisions &  55\\\hline
 Groundings & 44\\\hline
 Machinery failures & 21\\\hline
 Fire & 15\\\hline
 Total & 135\\\hline
\end{tabular}
\caption{Accident types and number of reports addressed in this study.}
\label{tab:3.1}
\end{center}
\end{table}

\subsection{Preprocessing}
A maritime accident investigation report is written in a natural language, by different investigating officers and hence does not follow a standard reporting format. This makes the investigation reports  inconsistent and noisy. If data is inconsistent, the text mining algorithms under-perform. The text data also contains some special formats like number formats, date formats and the most common words that are unlikely to help text mining such as prepositions, articles, and pronouns that are to be eliminated. In order to extract data which is consistent and accurate,  data preprocessing methods are  crucial. 

This section of the study reviews some simple NLP processing tasks that are used in the experiments, such as, tokenization and stemming using Natural Language Toolkit (NLTK). The NLTK, is a suite of Python libraries and programs for symbolic and statistical natural language processing~\cite{loper2002nltk, loper2004nltk}. NLTK includes graphical demonstrations and sample data. It is accompanied by extensive documentation. 

Some times the data is in Portable Document Format (PDF) and processing a PDF file is difficult. Hence, conversion  of data from PDF to TXT format is crucial.

\subsubsection{Tokenization}
The aim of the tokenization is to explore the words in a sentence~\cite{webster1992tokenization}. Textual data is only a block of characters at the beginning. But all the following processes in text classification require the words of the dataset. Hence, the tokenization is a pre-requisite
for data processing~\cite{mon2010myanmar}. 

This may sound trivial as the text is already stored in machine-readable formats. Nevertheless, some problems are still left, like the removal of punctuation marks. Other characters like brackets, hyphens, etc. require processing as well. Furthermore, the text should be lower cased to cater consistency in the documents. The main use of tokenization is identifying the meaningful significant words. Inconsistency can arise from different number formats or time formats. Another problem is abbreviations and acronyms which have to be transformed into a standard form.  

The following three-line program imports the \texttt{tokenize}
package, defines a text string, and then tokenizes the string on
whitespace to create a list of tokens.  Here "\url{>>>}" is
Python's interactive prompt; "\url{...}" is the second-level prompt.

{\small\begin{verbatim}
>>> from nltk_lite import tokenize
>>> text = 'Hello world.  This is a test.'
>>> list(tokenize.whitespace(text))
['Hello', 'world.', 'This', 'is', 'a', 'test']
\end{verbatim}}

\subsubsection{Stop Words}

\begin{figure}[htbp]

  \centering
  \includegraphics[width=3.5in, height=0.5in]{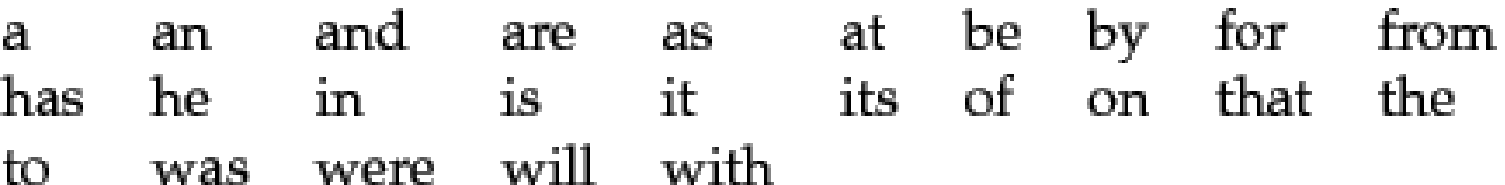}
  \caption[A list of 25 stop words]{\footnotesize  A stop word list of 25 semantically non-selective words which are common in Reuters-RCV1 dataset.}
  \label{fig:text.1}
\end{figure}

In text mining, most frequently used words or words that do not carry any information are known as
stopwords~\cite{manning1999foundations}. A example stoplist in English is shown in Figure~\ref{fig:text.1}. Typically a stop list constitutes about 400 to 500 such words and accounts for 20-30\% of the total word counts~\cite{yang1997comparative}. Hence, it  important to remove stopwords in improving the effectiveness and efficiency of an application. For an application, an additional domain specific stopwords list may be constructed~\cite{makrehchi2008automatic}. Most researches remove the stopwords using a standard stopword list.  An alternate way is to remove the most frequent words.
% figure 2.7 %
% end figure %

\subsubsection{Stemming}

Stemming refers to the process of reducing terms to their stems or root variants~\cite{paice1996method}. For example: 
\begin{itemize}
 \item agreed $->$ agree
\item meetings, meeting $->$ meet
\item  engineering, engineered, engineer $->$ engine
\end{itemize}

In statistical analysis, it helps greatly when comparing texts to identify words with a common meaning and form as being identical. For example, the words 'stopped' and 'stopping' stem from the same word 'stop'. Stemming identifies these common forms and reduces the computing time as different form of words is stemmed to form a single word. The most popular stemmer in English is Martin Porter's Stemming Algorithm~\cite{porter}, as shown to be effective in 
many cases~\cite{frakes1992stemming, perkins2010python, taghva2010effects}. 

 The following simple code demonstrates the stemming process using NLTK:
{\small\begin{verbatim}
>>> text = 'stemming can be fun and exciting'
>>> tokens = tokenize.whitespace(text)
>>> porter = tokenize.PorterStemmer()
>>> for token in tokens:
...     print porter.stem(token),
stem can be fun and excit
\end{verbatim}}

There are a few demerits of stemming. Firstly information about the full terms is lost. Secondly there is a trade-off between two main errors in stemming i.e 1) over-stemming and 2) under-stemming. Over-stemming occurs when two words with different stems are stemmed to the same root. This is also known as a false positive. Under-stemming happens when two words that should be stemmed to the same root are not. This is also known as a false negative. ~\cite{Paice:1990:STE:101306.101310, paice1994evaluation} showed that light-stemming reduces the over-stemming errors but increases the
under-stemming errors. On the other hand, heavy
stemmers reduce the under-stemming errors while
increasing the over-stemming errors.

\subsubsection{Zipf's Law}

Zipf’s law is the observation of~\cite{zipf19k} on the distribution of words in natural languages. It describes the word behavior in an entire corpus and can be regarded as a roughly accurate characterization of certain empirical facts. According to Zipf’s law, 

Frequency * rank = constant. 

Suppose $f(w)$ is the frequency of a word $w$ in free text.  Here, frequency is the number of times a word occurs in a corpus. If we compute the frequencies of the words in a corpus, and arrange them in decreasing order of frequency, then the product of the frequency of a word and its rank (its position in the list) is more or less equal to the product of the frequency and rank of another word. So frequency of a word is inversely proportional to its rank. That is, the frequency of words multiplied by their ranks in a large corpus is approximately constant. For example, the 50th most common word type should occur three times as frequently as the 150th most common word type. 

Researchers~\cite{wyllys1981empirical, fedorowicz1982theoretical, egghe1991exact, blair1990language} used the Zipf’s law to experiment on a large corpus. They found that only a small number of words occur more often than a large number of words that occur with low frequency. Between these two extremes there are medium frequency words as well. This distribution has its impact only on medium frequency words, having content-bearing terms. Common practice is to drop low frequency words as it has less discriminating power while the high frequency words are dropped using stop word list.

\subsubsection{Bag of Words Model}

The Bag of Words (BoW) model is a simplified text representation used in information retrieval (IR). In this model, a text is represented as an unordered collection of words, disregarding grammar and even word order. This model is commonly used in methods of document classification, where the occurrence of each word is used as a feature for training a classifier.

Text document representation based on the BoW model: 

Here are two simple text documents:
\begin{itemize}
\item   John likes to watch movies. Mary likes too.
\item   John also likes to watch football games.
\end{itemize}

Based on these two text documents, a dictionary is constructed as:
\begin{verbatim}
   {"John": 1, "likes": 2, "to": 3, 
   "watch": 4, "movies": 5, "also": 6, 
   "football": 7, "games": 8, "Mary": 9, 
   "too": 10}
\end{verbatim}
which has 10 distinct words. And using the indexes of the dictionary, each document is represented by a 10-entry vector:
\begin{verbatim}
   [1, 2, 1, 1, 1, 0, 0, 0, 1, 1]
   [1, 1, 1, 1, 0, 1, 1, 1, 0, 0]
\end{verbatim}

where each entry of the vectors refers to count of the corresponding entry in the dictionary. This vector representation does not preserve the order of the words in the original sentences. This study used Zipf's Law to obtain the dictionary, by removing the low frequency ($<5$) words to avoid a big feature space~\cite{manning1999foundations}.

\subsection{Document Representation}
\label{s:subjects}

A major challenge of the text classification problem is the representation of a
document. It is the final task in document preprocessing. The
documents are represented in terms of those features to which the dictionary was reduced
in the precedent steps. Thus, the representation of a document is a feature vector of $n$
elements where $n$ is the number of features remaining after finishing the selection process.   

When choosing a document representation, the goal is to choose the features that
allow document vectors belonging to different categories to occupy compact and disjoint
regions in the feature space~\cite{webb2003statistical}. There exist different types of information
that can be extracted from documents for representation. The simplest is the Bag-of-
Words representation (BoW) in which each unique word in the training corpus is used as
a term in the feature vector. Second type is the categorized proper names and named
entities (CAT) that only uses the tokens identified as proper names or named entities
from the training corpus used for representation~\cite{liddy2005improved}. 

A comprehensive study by \cite{basili2001hybrid} surveys the different
approaches in document representation that have been taken thus far and evaluates them
in standard text classification resources. The conclusion implies that more complex
features do not offer any gain when combined with state-of-the-art learning methods,
such as Support Vector Machines (SVM).

\subsubsection{Vector Space Model}

Vector Space Model (VSM) is a classical approach applied on text documents to obtain a matrix of numbers.  VSM has some severe drawbacks, resulting from its main assumption, reducing texts written in natural language, which is very flexible to strict mathematical representation. These problems, along with their possible solutions are discussed in this section.  

The vector space model is based on linear algebra and treats documents as
vectors of numbers, containing values corresponding to occurrence of words (also called terms) in respective documents~\cite{salton1975vector}. Let $t$ be size of the terms set, and $n$ be the size of
the documents set. Then, all documents $D_i$, $i = 1, \cdots , n$ may be represented as $t$-dimensional vectors:
\begin{equation} D_i = [a_{i1}, a_{i2}, \cdots, a_{it}] \end{equation}
where coefficients $a_{ik}$ represent the values of term $k$ in document $D_i$~\cite{salton1975vector}. Thus both documents and terms form a term-document matrix
$A_{(n \times t)}$. Rows of this matrix represent documents, and columns represent term vectors. Let
us assume that position $a_{ik}$ is set equal to 1, when term $k$ appears in document $i$, and to 0 when
it doesn't appear in it. For example, documents corresponding to a query "king", the corresponding term-document matrix can be created as shown in Table~\ref{tab:TDM}.  
\noindent Documents set:\\
$D_1$: The King University College\\
$D_2$: King College Site Contents\\
$D_3$: University of King College\\
$D_4$: King County Bar Association\\
$D_5$: King County Government Seattle Washington\\
$D_6$: Martin Luther King

\noindent Terms set $[T_1, T_2, \cdots , T_{15}]$: The, King, University, College, Site, Contents, of, County, Bar, Association,
Government, Seattle, Washington, Martin, Luther

\begin{table*}[t]
\begin{center}
\begin{tabular}{|c|c|c|c|c|c|c|c|c|c|c|c|c|c|c|c|}
\hline
 \tiny D/T & $T_1$ & $T_2$ & $T_3$ & $T_4$ & $T_5$ & $T_6$ & $T_7$ & $T_8$ & $T_9$ &  $T_{10}$ & $T_{11}$ & $T_{12}$ & $T_{13}$ & $T_{14}$ & $T_{15}$ \\
\hline
$D_1$ & 1 & 1 & 1 & 1 & 0 & 0 & 0 & 0 & 0 & 0 & 0 & 0 & 0 & 0 & 0\\
\hline
$D_2$ & 0 & 1 & 0 & 1 & 1 & 1 & 0 & 0 & 0 & 0 & 0 & 0 & 0 & 0 & 0\\ 
\hline
$D_3$ & 0 & 1 & 1 & 1 & 0 & 0 & 1 & 0 & 0 & 0 & 0 & 0 & 0 & 0 & 0\\ 
\hline
$D_4$ & 0 & 1 & 0 & 0 & 0 & 0 & 0 & 1 & 1 & 1 & 0 & 0 & 0 & 0 & 0\\ 
\hline
$D_5$ & 0 & 1 & 0 & 0 & 0 & 0 & 0 & 1 & 0 & 0 & 1 & 1 & 1 & 0 & 0\\
\hline
$D_6$ & 0 & 1 & 0 & 0 & 0 & 0 & 0 & 0 & 0 & 0 & 0 & 0 & 0 & 1 & 1\\ 
\hline 
\end{tabular}
\caption{Term-document matrix for an example document collection.}
\label{tab:TDM}
\end{center}
\end{table*}

\subsubsection{Merits and Demerits: VSM}
Using linear algebra as the basis of the vector space model is a merit. After transforming documents to vectors linear algebraic mathematical operations can be easily applied. Simple, efficient data structures may be used to store data. Representation of documents in the vector space model is very simple.  However, often these vectors are sparse, i.e. most of contained values are equal to 0. Hence, sparse vectors could be used to save memory and time.  

In basic vector space model, only occurrence of terms in documents is of importance and their order is not considered. It is the main reason why this
approach is often criticized~\cite{zamir1998web, weiss2001clustering}, as the information about the
proximity between words (their context in sentence) is not utilized. Consider for example two
documents: one containing a phrase "White House", which has a very specific meaning, and
another containing a sentence "A white car was parked near the house". Treating documents
simply as sets of terms we only know that words "white" and "house" occur in both
documents, although their context there is completely different.
However, this problem can be easily overcome - one can supplement this model, using
also phrases in addition to terms in document vectors, as described in~\cite{maarek1991information} and~\cite{maarek2000ephemeral}.  

\subsection{Term Weights}

The process of calculating weights of terms is called terms weighting. It addresses how important a term is with respect to a document (since not all terms are equally informative about the contents of the document). According to ~\cite{how2004empirical},  term weighting is used to describe and summarize document content based on a term's importance. There are several main methods used to assign weights to terms. The simplest method is boolean terms weighting, which, as its name suggests, sets weights to 0 or 1 depending on the presence of term in document. This method is used to
calculate the term-document matrix in example shown in Table~\ref{tab:TDM}. Using this method
causes loss of valuable information, as it differentiates only between two cases: presence or
absence of term in document, and exact number of occurrences of word may indicate its
importance in documents.  

The method utilizing knowledge of exact number of term occurrences in documents is called TF term weighting (TF stands for Term Frequency). TF is the total count of the particular word repeated in a document and is calculated as \begin{equation}tf_{ij} = \frac{n_{i,j}}{\sum\limits_k n_{k,j} }\end{equation} where $n_{i,j}$ is the number of times the term $t_i$ occurs in document $d_j$ and the denominator is the sum of number of times all terms occur in document $d_j$~\cite{manning1999foundations}. 

Document Frequency (DF) is defined as the total number of times the word occurs in the entire documents i.e. number of
documents containing the significant word. On the other hand, Inverse Document Frequency (IDF) is a measure of whether the term is common or rare across all documents~\cite{robertson2004understanding}. It is obtained by dividing the total number of documents by the number of documents containing the term, and then taking the logarithm of that quotient.

\begin{equation}
idf_{i} = \log \frac{|D|} {|d:t_i \in d|},
\label{eq1}
\end{equation} 
here $|D|$ is the total number of documents in the collection and $|d:t_i \in d|$ is the number of documents where the term $t_i$ appears~\cite{manning1999foundations}.

TFIDF has three assumptions that, in one form or another, will appear practically in all weighting methods:

\begin{itemize}
\item IDF assumption : "rare terms are no less important than frequent terms".
\item TF assumption : "multiple appearances of a term in a document are no less
important than single appearances". 
\item Normalization assumption :  "for the same quantity of term matching, long documents are no more important than short documents".  
\end{itemize}

A classical term weighting method that takes into account both term and document
frequencies is called tf-idf terms weighting, and is probably the most popular approach in
information retrieval systems~\cite{paukkeri2011effect, taghva2010effects}. Term weight in this method is calculated as a product of its
term and inverse document frequencies, hence its name.
\begin{equation}
TF-IDF_{ij} = {tf_{i,j} \times idf_{i}}
\end{equation}

\section{Experiments and Results}
This chapter describes the procedure and results for extracting the causal relations using pattern classification and connectives methods. 

\subsection{Pattern Classification Method}
The implementation of pattern classification method and its results are discussed in this section.
\subsubsection{Dataset Collection}

The dataset is the collection of causal relations marked by three domain experts. The experts marked a total of 151 causal sentences in four accident investigation reports. These 151 causal sentences with an addition of 151 non-causal sentences from the same accident investigation reports are combined to form a complete dataset containing 302 sentences. Out of them, 70\% i.e 212 sentences(106 causal and 106 non-causal) are considered as the training set and remaining 30\% i.e 90
sentences (45 causal and 45 non-causal) are considered to be test set.

\subsubsection{Data Preprocessing}

The documents collected are converted from PDF to TXT format. The training data in TXT format needs to be tokenized as explained in  Chapter 3, after which the stop words are removed. The list \footnote{\url{http://www.ranks.nl/resources/stopwords.html}} of stop words used in the study is 416 words including single characters and excluding the transitions, conjunctions and verbphrases listed in section 2.1. Before removing stop words the total number of terms from 212 sentences is 6790. After removing stop words, remaining number of terms are 3414. In the next step, stemming is performed and unique words are recorded. Words that occur 5 times or less are also removed in this process. Finally we are left with a list of \textit{significant words}. Hence our final features are a total of 990 significant terms. 

\subsubsection{Data Representation}

Using these 990 significant terms a train:document-term matrices are constructed based on TF and TFIDF weights. Similarly test dataset is tokenized and stemmed using Porter's stemmer. Based on significant terms collected from the training set, the test:document-term matrices are constructed for both TF and TFIDF weights.

\subsubsection{Classifiers}
The train:document-term matrices for both the weighting schemes are divided into 10 folds, where each fold consists of 124 samples as training set and 14 samples as validation set (few folds included 125:13 samples). Each fold is given as an input to the classifier algorithms viz. 1) Naive Bayes classifier 2) SVM-Linear kernel classifier 3) SVM-Gaussian kernel classifier. Naïve Bayes classifier is based on Multinomial distribution\footnote{\url{http://www.mathworks.se/help/stats/naivebayes.fit.html}}, which is used for classifying the count-based data such as the Bag of Words (BoW) model. 

\subsubsection{Parameter Tuning}
SVM-Linear kernel classifier is used with a near boundary coefficient value = 10 and SVM-Gaussian kernel classifier is used with a sigma value = 16. These values were considered after running the classifier with $C\;=\;\{0.01, 0.1, 1, 10, 100\}$ and $\;sigma\;=\;\{8, 16, 32, 64, 128\} $. 

Table~\ref{tab:res.1} shows the F-measure on the validation set for SVM-Gaussian classifier for various sigma values ($sigma\;=\;\{8, 16, 32, 64, 128\}$). It can be seen that the performance is best when sigma value is 16.

\begin{table}[h]
\centering
\begin{tabular}{|c|c|c|}
\hline
 Sigma & F-Measure (TF) & F-Measure (TFIDF)\\
\hline
  8 &      0.6885 & 0.5370\\
 \textbf{16} &      \textbf{0.7826} &      \textbf{0.6716}\\
  32 &      0.6817 &  0.6667\\
  64 &      0.0000 &  0.6667 \\
  128 &      0.0000 & 0.0000 \\
\hline
\end{tabular} 
\caption{Parameter tuning for SVM-Gaussian Kernel Classifier}
\label{tab:res.1}
\end{table}

\noindent Table~\ref{tab:res.2} shows the F-measure on the validation set for SVM-Linear classifier for various C values ($C\;=\;\{0.001, 0.01, 0.1, 1, 10\})$. It could be seen that the performance of the Linear kernel is best when $C$ value is 10.

\begin{table}[h]
\centering
\begin{tabular}{|c|c|c|}
\hline
 C & F-Measure (TF) & F-measure (TF-IDF)\\
\hline
  0.01 &      0.5000 &  0.5000\\
  0.1 &      0.5000 &  0.5094 \\
  1 &      0.5566 &  0.5377\\
 \textbf{10} &   \textbf{0.7604} &  \textbf{0.7217}\\
  100 &      0.6132 &  0.6085\\
\hline
\end{tabular} 
\caption{Parameter tuning for SVM-Linear Kernel Classifier }
\label{tab:res.2}
\end{table}

\subsubsection{Cross Validation and Testing}
\noindent Table~\ref{tab:res.3} depicts the 10 fold cross validation on various classifiers used in this experiment on TF weighting scheme. The results of the 10 fold cross validation is evaluated against F-measure. Naïve Bayes classifier achieved 71\% F-measure across all the folds. SVM classifier with Gaussian and Linear kernels out performed with 74\% and 73\% F-measure respectively.

\begin{table}[h]
\centering
\begin{tabular}{|c|c|c|c|}
\hline
Fold & Naïve Bayes & SVM-Linear & SVM-Gaussian\\
\hline
1	&	0.6667	&	0.8696	&	0.8333	\\
2	&	0.6	&	0.7273	&	0.64	\\
3	&	0.9231	&	0.8	&	0.8	\\
4	&	0.8148	&	0.7407	&	0.8148	\\
5	&	0.9167	&	0.88	&	0.9167	\\
6	&	0.7826	&	0.6667	&	0.6667	\\
7	&	0.8276	&	0.8	&	0.8462	\\
8	&	0.7368	&	0.8	&	0.7619	\\
9	&	0.3158	&	0.5	&	0.56	\\
10	&	0.5	&	0.6	&	0.5714	\\
\hline
\hline
\textbf{Average} &      \textbf{0.7084}   &       \textbf{0.7384}  &        \textbf{0.7411} \\
\hline
\end{tabular} 
\caption{F-measure on validation sets for TF weighting scheme.}
\label{tab:res.3}
\end{table}

\noindent Table~\ref{tab:res.4} depicts the 10 fold cross validation on various classifiers used in this experiment for TFIDF weighting scheme. It can be seen that Naive Bayes and SVM-Gaussian classifiers achieved 69\% F-measure on an average, where as SVM-Linear classifier achieved only 46\% F-measure.

\begin{table}[h]
\centering
\begin{tabular}{|c|c|c|c|}
\hline
Fold & Naïve Bayes & SVM-Linear & SVM-Gaussian\\
\hline
1	&	0.625	&	0.3158	&	0.4444	\\
2	&	0.7857	&	0.5455	&	0.8108	\\
3	&	0.5455	&	0.4706	&	0.6667	\\
4	&	0.7143	&	0.3529	&	0.7586	\\
5	&	0.75	&	0.5556	&	0.6	\\
6	&	0.8	&	0.25	&	0.8333	\\
7	&	0.6667	&	0.7143	&	0.6154	\\
8	&	0.9167	&	0.6364	&	0.9091	\\
9	&	0.6667	&	0.4	&	0.8	\\
10	&	0.4762	&	0.375	&	0.5556	\\
\hline         
\hline
\textbf{Average} &      \textbf{0.6947}   &       \textbf{0.4616}  &        \textbf{0.6994} \\
\hline
\end{tabular} 
\caption{F-measure on validation sets for TFIDF weights}
\label{tab:res.4}
\end{table}

\noindent Table~\ref{tab:res.5} illustrates the performance of the test set on both TF and TFIDF weighting scheme. It is clearly seen that the SVM classifiers have achieved almost 70\% of F-measure on TF weights.

\begin{table}[h]
\centering
\begin{tabular}{|c|c|c|c|}
\hline
Weights & Naive Bayes & SVM-Linear & SVM-Gaussian\\
\hline
TF &    0.5882  &  0.6916 &    0.7207 \\
TFIDF & 0.4941  &  0.3143 &    0.5825 \\
\hline
\end{tabular} 
\caption{F-measure on test set}
\label{tab:res.5}
\end{table}

\begin{figure}[h]

  \centering
  \includegraphics[width=1\linewidth]{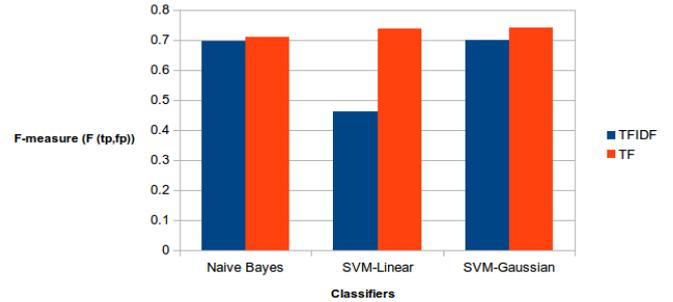}
  \caption{Comparison of F-measure ($F_{tp,fp}$) on validation sets for TF \& TFIDF weights.}
  \label{fig:res.1}
\end{figure}

\noindent Figure~\ref{fig:res.1} compares F-measure ($F_{tp,fp}$) on K-fold cross validation sets for both TF and TFIDF weighting schemes. It is clear that there is a marginal increase in the F-measure on the performance of naïve Bayes and SVM-Gaussian classifier when using TF weights. SVM-Linear classifier showed an increase of 27\% when using TF weights.

\begin{figure}[h]

  \centering
  \includegraphics[width=1\linewidth]{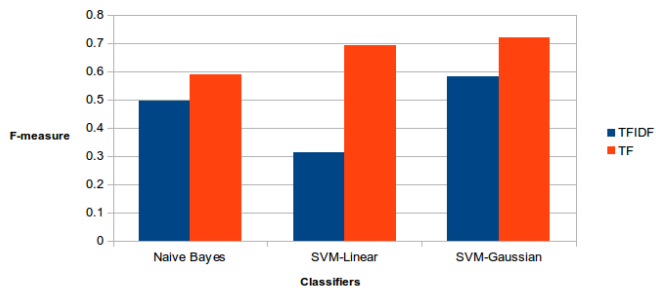}
  \caption{Comparison of F-measure on test sets for TF \& TFIDF weights.}
  \label{fig:res.2}
\end{figure}

\noindent Figure~\ref{fig:res.2} illustrates the comparison of F-measure on test-sets for both TF and TFIDF weighting schemes. There is a 10\% increase in the F-measure on the performance of naïve Bayes and SVM-Gaussian classifier when using TF weights, while SVM-Linear classifier showed a significant increase of 38\% when using TF weights. 

To summarize, all the three classifiers have achieved more than 70\% F-measure across the folds using TF weights. When using TFIDF weights, naïve Bayes and SVM-Gaussian classifiers have achieved 69\% F-measure across the folds. SVM-Linear classifier achieved only 46\% F-measure. A marginal increase in the F-measure is recorded on the performance of naïve Bayes and SVM-Gaussian classifier when using TF weights. Performance on the test set illustrates that the SVM classifiers have achieved almost 70\% of F-measure on TF weights.
These results clearly show that the weighting scheme TF outperforms TFIDF. The possible reason for this phenomenon is discussed in section~\ref{discuss}.

\subsection{Connectives Method}

In this section, the implementation procedure for connectives method is discussed along with the results.

\subsubsection{Dataset Collection}

The dataset is a collection of four accident investigation reports, where each report is marked by three domain experts for causal relations. A total of 151 causal relations are marked.

\subsubsection{Implementation}

From the dataset, the sentences which have connective words such as, transitions, conjunctions and verb
phrases described in section~\ref{s:conmen} and listed in Table~\ref{tab:1}, Table~\ref{tab:2}, Table~\ref{tab:3} and Table~\ref{tab:4} are extracted using linux command \textit{grep}\footnote{\\url{http://linux.die.net/man/1/grep}} and then collected to a new file. A MAIB report typically consists of 60 pages and the causal relations extracted from an accident investigation report averages on 10 sentences. Hence a 60 page report is transformed to a half page text document including major contributory causes. Some example causal relations extracted from few reports, are as follows:

\begin{itemize}
\item Cause 1: "In assessing that Boxford was overtaking the 
fishing vessel, it is clear that the master misinterpreted the lights he saw. \textit{Consequently}, his alteration to starboard to keep clear of Admiral Blake only served to reduce an already small CPA, thereby exacerbating the close-quarters situation."

\item Cause 2: "The master did not activate Saffier's general alarm or alert the crew in any other way. \textit{Consequently} they had limited warning to prepare for, or react to, the subsequent damage."

\item Cause 3: "No fire detection or fire suppression systems were fitted.  \textit{As a result}, the fire was able to develop undetected for about  minutes."

\item  Cause 4: "The distortion and subsequent cracking of the furnace tube in the auxiliary boiler was \textit{due to} sustained overheating."

\item Cause 5: "The scenario that the fire was \textit{caused} when hot debris from the hotwork on the hopper came into contact with the conveyor belt."

\item Cause 6: "Actions to reduce, or stop, the sheer, were insufficient to counteract the forces acting on the hull. \textit{Therefore}, control of Arold was lost and a collision with the approaching Anjola ensued."

\end{itemize}

\subsubsection{Exploratory analysis}

Instead of reading a whole investigation report, one could read the extracted causal relations from the investigation report to find out the contributory causes for the marine accident. The causal relations extracted from an investigation report are shown below: 

\begin{itemize}
\item  In assessing that Boxford was overtaking the fishing vessel, it is clear that the master misinterpreted the lights he saw.  Consequently, his alteration to starboard to keep clear of Admiral Blake only served to reduce an already small CPA, thereby exacerbating the close-quarters situation.
\item However, these criticisms were at variance with the radar's performance log that indicated the S-band radar was functioning correctly. Therefore, it is equally likely that the failure to detect Admiral Blake by radar was due to the radar's settings not being optimized for the prevailing sea state and the range scale selected. 
\item However, the deck cadet on Boxford did not report the fishing vessel's lights until she was at about nm ahead.
 This was probably because the fishing vessel's lights were only intermittently visible due to Admiral Blake's  MV Boxford's view ahead partially obstructed by the uprights of the deck cranes,  Boxford's master was unable to detect Admiral Blake by radar.

\end{itemize}

From these causal relations, it is clearly seen that the contributory causes for the accident, was the radar's performance not being optimized for the prevailing sea state.

\subsubsection{Evaluation}

The evaluation is subjective since experts have marked the causal sentences according to their subjective views. In this kind of situation, sometimes qualitative evaluation outweighs the quantitative. To qualitatively evaluate the performance of the connectives method (automatic algorithm), a questionnaire is given to the domain experts. The questionnaire and experts' answers are shown in  Table~\ref{tab:res1}. For quantitative evaluation, Precision and Recall from the context of IR model is adapted. Here, retrieved are the sentences that algorithm marked as causal, denoted by 'A'. Relevant are the ones that experts' have marked as causal sentences, denoted by 'E'. The Precision is evaluated as $(E \cap A) / A$ and Recall is given by $(E \cap A) / E$. F-measure is evaluated as $(2 \times P \times R) / (P+R)$.

\begin{table*}
\centering
\begin{tabular}{|p{3cm}|p{3.5cm}|p{3.5cm}|p{3cm}|}
\hline
 Question & Expert-1 & Expert-2  & Expert-3\\
\hline
\footnotesize In what kind of situations do the automatic algorithm and the expert agree? or do not agree?, if so what are they ?  &     \footnotesize They agree on many of the sentences, but the expert has also considered many more passages of text as causal information. Further, they especially
disagree on safety management related text.
 & \footnotesize     It agrees in most cases. Especially the algorithm extracted the causal chains pertaining to the accidents, which is of the expert's Interest. & \footnotesize The automatic algorithm and expert agree for important causes behind the accidents. They do not agree for safety policy information since that information does not have causal information in them, yet they are important in expert’s point of view. \\
\hline

\hline
\footnotesize What does the algorithm find that the expert didn't consider? & \footnotesize     Basically the algorithm extracts longer fractions of the text and also some redundant information that was already found before in other part of the report (expert had marked them only once) &    \footnotesize The algorithm found almost what expert has considered and also some extra information but always contextual information is needed. & \footnotesize The algorithm found much more information than what expert had marked. Expert agree that the information marked by the algorithm is important. \\
\hline

\hline
\footnotesize What kind of sentences/ expression/ information the expert found in the automatic causal relations extraction? What are expert's generalizations about them?
  &   \footnotesize   The information the algorithm found was almost always clearly stated sentences of the investigator’s reasoning what might have been causing the accidents. The algorithm seems to find these quite well.

 &  \footnotesize    The algorithm found the causal chains very well. Before reading a whole report, this algorithm could  be employed to capture causal chains, which could make reading more easier. &  \footnotesize Useful and important causal information leading to the accidents was found in the automatic causal relations extracted. It would also be interesting to see the algorithm extracting the information related to safety policies. \\
\hline

\hline
\footnotesize What had the expert considered important but the algorithm did not find?
  &   \footnotesize   Safety management related information, sentences which described various situational factors related to the accident but which were not mentioned within a clear causal sentence form. & \footnotesize    Very few sentences were missed out by the algorithm, but it works reasonably well when extracting the automatic causal chains. & \footnotesize Expert considered few safety policies to be important which algorithm did not find. But expert understands that those sentences are not accurately causal.  \\
\hline
\end{tabular} 
\caption{Questionnaire and experts' answers.}
\label{tab:res1}
\end{table*}

Expert-1 agrees that the algorithm performs well but some passages contain non causal information and does not sufficiently represent the safety management related text. She also noted that the algorithm extracted longer fractions of the text and marked some redundant information. According to her, the algorithm found clearly stated sentences of the accident causes. But the sentences describing various situational factors to the accident were not mentioned within a clear causal sentence format. Table~\ref{tab:res2} shows a total number of 110 causal sentences marked by expert-1 which are relevant. The average values of precision, recall and F-measure for connectives method on expert-1 marked reports are 0.60, 0.54 and 0.57 respectively. 

\begin{table*}[t]
\centering
\begin{tabular}{|c|c|c|c|c|c|c|}
\hline													
Report	&	E	&	A	&	$E \cap A$	&	$(E \cap A) / A$	&	$(E \cap A) / E$	&	F-measure	\\
\hline													
1	&	32	&	26	&	13	&	0.5	&	0.41	&	0.45	\\
2	&	29	&	27	&	16	&	0.59	&	0.55	&	0.57	\\
3	&	16	&	15	&	9	&	0.6	&	0.56	&	0.58	\\
4	&	33	&	30	&	21	&	0.7	&	0.64	&	0.67	\\
\hline
Total	&	110	&	98	&	59	&	mean=0.6	&	mean=0.54	&	mean=0.57	\\
\hline					
\end{tabular} 
\caption{Performance of connectives method on expert-1 marked reports.}
\label{tab:res2}
\end{table*}

The second expert found interesting information which the algorithm unearthed pertaining to the accidents. The algorithm performed as per his expectations although in some instances context was needed. He claimed that it would be easier to read the report generated by the algorithm to capture essential information. Table~\ref{tab:res3} shows a total number of 110 causal sentences marked by expert-2. The total number of causal sentences both expert-2 and algorithm agree on is 61. The average values of Precision, Recall and F-measure are 0.62, 0.75 and 0.68 respectively.  

\begin{table}[h]												
\centering												
\begin{tabular}{|c|c|c|c|c|c|c|}												
\hline												
Report	&	E	&	A	&	$E \cap A$	&	$(E \cap A) / A$	&	$(E \cap A) / E$	&	F-measure \\
\hline												
1	&	20	&	26	&	17	&	0.65	&	0.85	&	0.74\\
2	&	22	&	27	&	19	&	0.7	&	0.86	&	0.77\\
3	&	12	&	15	&	9	&	0.6	&	0.75	&	0.67\\
4	&	27	&	30	&	16	&	0.53	&	0.59	&	0.56\\
\hline
Total	&	81	&	98	&	61	&	mean=0.62	&	mean=0.75	&	mean=0.68\\
\hline												
\end{tabular} 
\caption{Performance of connectives method on expert-2 marked reports.}
\label{tab:res3}
\end{table}

The expert-3 reiterated the views expressed by expert-1 in stating that the algorithm missed out some safety policy information. He stated that the algorithm performed better than expected in mining automatic casual information. Table~\ref{tab:res4} shows a total number of 60 causal sentences marked by expert-3 (relevant). The total number of causal sentences that both expert-3 and algorithm agree on is 40. The average values of Precision, Recall and F-measure are 0.41, 0.67 and 0.51 respectively.

\begin{table}[h]   												
\centering												
\begin{tabular}{|c|c|c|c|c|c|c|}
\hline													
Report	& E & A & $E \cap A$	& $(E \cap A) / A$ & $(E \cap A) / E$ & F-measure\\
\hline													
1	&	17	&	26	&	14	&	0.54	&	0.82	&	0.65	\\
2	&	11	&	27	&	7	&	0.26	&	0.64	&	0.37	\\
3	&	7	&	15	&	5	&	0.33	&	0.71	&	0.45	\\
4	&	25	&	30	&	14	&	0.47	&	0.56	&	0.51	\\
\hline
Total	&	60	&	98	&	40	&	mean=0.41	&	mean=0.67	&	mean=0.51	\\
\hline													
\end{tabular} 
\caption{Performance of connectives method on expert-3 marked reports.}
\label{tab:res4}
\end{table}

\begin{figure}[h]

  \centering
  \includegraphics[width=1.05\linewidth]{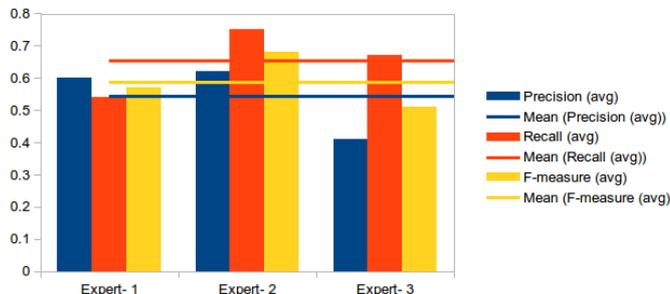}
   \caption[Comparison of evaluation metrics for connectives method.]{\footnotesize Comparison of Precison, Recall and F-measure for connectives method on the experts' marked documents.}
  \label{fig:res.3}
\end{figure}

To summarize, all the experts expressed the opinion that the algorithm performed reasonably well. When it comes to bringing safety policies to light it could be improved. Figure~\ref{fig:res.3} shows that connectives method gave a good performance on expert-2 marked documents. F-measure on expert-2 marked reports is 68\% and is greater in comparison with expert-1 (57\%) and expert-3 (51\%). The average value of F-measure on connectives method is 58\%.  

\section{Conclusions \& Discussions}

The objective of this study is to extract causal relations from maritime accident investigation reports. The data used in this study was a collection of 302 sentences (151 causal and 151 non-causal sentences). The training and test set consisted of 212 (106 causal and 106 non-causal) and 90 sentences (45 causal and 45 non-causal) respectively. To achieve the objective, this study presented two schemes of extraction techniques, namely : 1) Pattern classification method and 2) Connectives method.

Pattern classification method used naïve Bayes and Support Vector Machines (SVM) as classifiers. The input to the classifiers were the document-term matrices, where documents represented the causal and non-causal sentences and the terms represented the Bag of Words (BoW). The document-term matrices were constructed using both TF and TFIDF weighting schemes. The naïve Bayes classifier considered multinomial distribution and SVM classifiers used Linear and Gaussian kernels. For the latter classifier, parameter tuning was performed to obtain optimal parameters holding best for the classification results.

The K-fold cross validation on all the three classifiers achieved more than 70\% F-measure on an average using TF weights. When using TFIDF weighting scheme, naïve Bayes and SVM-Gaussian classifiers achieved 69\% F-measure across the folds, while SVM-Linear classifier achieved only 46\% F-measure. A marginal increase in the F-measure was recorded on the performance of naïve Bayes and SVM-Gaussian classifier when using TF weights. Performance on the test set illustrates that the SVM classifiers have achieved almost 70\% of F-measure on TF weights. 

Connectives method of implementation was rather simpler. A linux command 'grep' extracted all the causal relations based on connective words listed in this study. The F-measure recorded on expert-1 and expert-3 marked reports are 57\% and 51\% respectively. The F-measure on expert-2 marked reports is high with 68\%. Hence this study shows that, using text mining methods, the causal patterns can be fairly extracted from marine accident investigation reports with a reasonable F-measure. Comparing the pattern classification method (F-measure (average: 65\%)) with connectives method F-measure (average: 58\%), shows pattern classification method gave a fair and sensible performance.  

\subsection{Discussion}
\label{discuss}
The results on the test set clearly show that the weighting scheme TF outperforms TFIDF. A high weight in TFIDF is reached by a high term frequency (in the given document) and a low document frequency of the term in the whole collection of documents. Hence TFIDF weights tend to filter out common terms. Since the ratio inside the IDF's log function is always greater than or equal to 1, the value of IDF (and TFIDF) is greater than or equal to 0. As a term appears in more documents, the ratio inside the logarithm approaches 1, bringing the IDF and TFIDF closer to 0. This study included most common terms such as: transition words, conjunction words and causal verb phrases (chapter 2, section 2.3). These words were influential on the performance of classifiers using TFIDF weights.

Machine learning studies for example in~\cite{baldi2000assessing} reveal that, if the datasets used for training and testing a particular classification algorithm are very similar, the apparent predictive models' performance may be overestimated, reflecting the ability of the model to reproduce its input rather than its ability to interpolate and extrapolate. Hence, the actual level of prediction accuracy depends on the degree of similarity between training and test datasets, which can explain the performance of different classifiers being relatively constant with the amount of training data.

The dataset contained 151 data samples corresponding to each class. In such a case, 70\% of data, i.e. 212 data points were used for training the classifiers. With such a small amount of training data, SVM classifiers generally generate an over-fit or under-fit learning model. Moreover, with lower amounts of training data, naïve Bayes which is expected to show better performance failed to reach the average classification accuracies obtained by SVM. Similarly, in the case where 90\% training data and 10\% data were used for validating, naïve Bayes failed to compete with SVM learning models (as shown in Table~\ref{tab:res.3} and Table~\ref{tab:res.4}). A possible reason for such behavior of naïve Bayes classifier can be explained by redundancy in the data used for training and validating the classifiers \cite{baldi2000assessing}. 

The most important limitation concerning the implementation of this study is the lack of labeled data. Though there were 135 accident investigation reports, the analysis considered only 4 reports that have been marked by the experts. It is still unclear if one can address the "ground truth" of the expert's marked sentences as the truth. The labeled data is subjective and necessarily one can not say much about the performance of the methods employed in the study as the evaluation is subjective. In this kind of situations sometimes qualitative evaluation outweighs the quantitative. There also arises a question whether the evaluation based on these facts is reliable as such. Nevertheless, it plays a crucial role in the performance of the algorithms.  

To conclude, it is possible to say that experts' marked causal relations from four different accident investigation reports were fairly sufficient to classify and extract causal patterns from other accident investigation reports. The results also suggest that usage of connecting words were influential on classification results. It was evident from this analysis, that pattern classification method outweighs the connectives method. It is still unclear which of the approaches are most suitable for exacting causal relations from maritime accident reports. When there are many similar methods available it is difficult to choose which one to use. In such a case simplicity and reputation of the method and experience of its usage can influence the decision. This research might embark on developing effective tools and methodologies in future for identifying human and organizational factors present in the accident investigation reports. 


%%%%%%%%%%%%%%%%%%%%%%%%%%%%%%%%%%%%%%%%%%%%%%%%%%%%%%%%%%%%%%%%%%%%%%%%%%%%%%%%

%%%%%%%%%%%%%%%%%%%%%%%%%%%%%%%%%%%%%%%%%%%%%%%%%%%%%%%%%%%%%%%%%%%%%%%%%%%%%%%%

%%%%%%%%%%%%%%%%%%%%%%%%%%%%%%%%%%%%%%%%%%%%%%%%%%%%%%%%%%%%%%%%%%%%%%%%%%%%%%%%
%\small{
\section*{ACKNOWLEDGMENT}
I thank CAFE project for funding this work in the year 2013.  CAFE project was financed by the European Union - European Regional Development Fund - Regional Council of Päijät-Häme, the City of Kotka, Kotka-Hamina regional development company Cursor Ltd., Kotka Maritime Research Association Merikotka and the following members of the Kotka Maritime Research Centre Corporate Group: Port of Hamina Kotka, Port of Helsinki,  Aker Arctic Technology Inc.  and Arctia Shipping Ltd. 

I would like to thank Professor Erkki Oja, Tiina Lindh-Knuutila and Maria Hänninen, who have guided me in this study in the year 2013. It is because of their support I was able to finish this study.

%%%%%%%%%%%%%%%%%%%%%%%%%%%%%%%%%%%%%%%%%%%%%%%%%%%%%%%%%%%%%%%%%%%%%%%%%%%%%%%%

\bibliographystyle{ieee}
\small{
\bibliography{citations,santosh_citations}
}

\end{document}